\documentclass{raa}
\usepackage{color}
\usepackage{graphicx,times}
\usepackage{natbib}
\usepackage{amssymb,amsmath}
\usepackage{threeparttablex}
\usepackage{multirow}
\usepackage{makecell}
\bibpunct{(}{)}{;}{a}{}{,}

\usepackage[a4paper=true,pagebackref=true]{hyperref}
\hypersetup{pdftitle = The title of my PDF, pdfauthor = My name, pdfsubject= The subject, pdfkeywords = keyword1 keyword2 keyword3}
\hypersetup{colorlinks = true, linkcolor = green, anchorcolor = red, citecolor = blue, filecolor = red, pagecolor = red, urlcolor = red}

\begin{document}

   \title{Pulsar discovery prospect of FASTA
}

 \volnopage{ {\bf 20XX} Vol.\ {\bf X} No. {\bf XX}, 000--000}
   \setcounter{page}{1}

   \author{Mengyao Xue\inst{1}, Weiwei Zhu\inst{1,2}, Xiangping Wu\inst{1,2}, Renxin Xu\inst{3}, Hongguang Wang\inst{4,5}
   }

   \institute{National Astronomical Observatories, Chinese Academy of Sciences, Beijing 100101, China; {\it mengyaoxue@nao.cas.cn}\\
        \and
             Institute for Frontiers in Astronomy and Astrophysics, Beijing Normal University,  Beijing 102206, China\\
        \and
             School of Physics, Peking University, Beijing 100871, China\\
        \and
             Department of Astronomy, School of Physics and Materials Science, Guangzhou University, Guangzhou 510006, China\\
        \and
             Great Bay Center, National Astronomical Data Center, Guangzhou, Guangdong 510006, China\\
\vs \no
   {\small Received 2023 Feb 6th; accepted 2023 Jun 2nd}
}

\abstract{The Five-hundred-meter Aperture Spherical radio Telescope (FAST) has discovered more than 650 new pulsars, which account for 20\% of our known Galactic pulsar population.
In this paper, we estimate the prospect of a pulsar survey with a radio telescope array to be planned --- the FAST Array (FASTA), consists of six "FAST-type" telescopes.
Such a sensitive radio telescope array would be a powerful instrument in probing the pulsar population deep into our Galaxy as well as in nearby galaxies. We simulate the FASTA pulsar discovery prospects with different Galactic pulsar population models and instrumental parameter combinations.
We find that FASTA could detect tens of thousands of canonical pulsars and well-over thousands of millisecond pulsars. We also estimate the potential yield if the FASTA is used to search for pulsars from the nearby spiral galaxy M31, and find that it would probably discover around a hundred new radio pulsars.
}

   \authorrunning{Xue et al. }            
   \titlerunning{Pulsar discovery prospect of FASTA}  
   \maketitle

%
\section{Introduction}           
\label{sec:intro}

Pulsars are rapidly rotating neutron stars. So far, astronomers have discovered more than 3300 radio pulsars\footnote{ATNF Pulsar Database v1.68; www.atnf.csiro.au/research/pulsar/psrcat} \citep{Manchester2005}.
The studies of pulsars have led to a number of important progress in the frontiers of astrophysics and physics, including the theory of gravitation, the stellar evolution, the matter distribution of our Galaxy, and the equation of state for ultra-dense matter.
The extraordinary impact and prominent scientific applications of pulsar research have made it one of the key science projects for nearly every new generation of advanced radio telescopes.

There are two main types of radio pulsars: canonical pulsars (CPs) and millisecond pulsars (MSPs). CPs are generally thought to be the remaining core after the supernovae explosion of progenitor massive star ($\sim$ 10 to 30 $M_{\odot}$). The radio emission of CPs are powered by the pulsar's rotation energy. Therefore, as CPs keep emitting radio signals, they will gradually spin down when they become older. For example, one of the youngest pulsars, the Crab pulsar, has a spin period of $\sim$33\,ms. Old CPs can have a spin period as long as tens of seconds \citep[e.g.][]{Tan2018,Caleb2022}. MSPs \citep{Backer1982} are old, rapidly rotating neutron stars that have been `spun up' or `recycled' through the accretion of matter from a companion star in a close binary system. Their typical spin period is in the range of about $\sim$1-30 milliseconds.

Pulsar surveys and the resulting pulsar discoveries are crucial resources for us to build up our understanding of the Galactic pulsar population. For example, the Parkes Multi-beam Pulsar Survey \citep[PMPS, ][]{Manchester2001}, which has discovered $\sim$800 pulsars initially \citep{Manchester2001,Morris2002,Kramer2003,Hobbs2004,Faulkner2004} and 1038 after many passes of searches \citep[e.g.][]{Eatough2013,Knispel2013,Bates2013}, provides a large sample of pulsar detections with uniform instrument settings. Using this sample, astronomers established basic models of the Galactic pulsar population, including their spatial, period, and luminosity distributions \citep{Lorimer2006,FK06}. \cite{Lorimer2006} suggest there are $(3.0\pm0.1)\times10^{4}$ canonical pulsars beaming toward the Earth with luminosities above 0.1\,mJy\,kpc$^2$, while \cite{FK06} predict there are $(1.2\pm0.2)\times10^{5}$ detectable canonical pulsars in the Galaxy. However, some following pulsar surveys using other instruments such as Arecibo PALFA \citep{Cordes2006}, Green Bank North Celestial Cap pulsar survey \citep[GBNCC;][]{Stovall2014}, and FAST Galactic Plane Pulsar Snapshot survey \citep[GPPS;][]{Han2021} suggest that the population extrapolated from PMPS might be overestimated. A possible explanation is  that PMPS concentrates on the Galactic Plane  which has a high density of pulsars \citep{Swiggum2014,GBNCC2020,Han2021}.

In the early 1990s, as one of the early concepts of the Square Kilometre Array (SKA),  Chinese radio astronomers proposed to build $\sim$\,30 large spherical reflectors in Guizhou, southwest China, with a diameter of roughly 200 to 300\,m each. This design was referred to as the Kilometer-square Area Radio Synthesis Telescope \citep[KARST; ][]{PengNan1997,Nan2002}. Although it was not selected as the final design for the SKA, it led to the birth of Five hundred meter Aperture Spherical Telescope \citep[FAST; ][]{Nan2006,NanLi2011,Li2018}, currently the most sensitive single-dish radio telescope. Since the commissioning operation of FAST, a number of notable science outcomes in pulsar studies have been carried out \citep[e.g. ][etc.]{Qian2020,Cameron2020,Han2021}.

Today, the plan of building large spherical reflectors array has once again been put on the agenda after the successful delivery and operation of FAST. Chinese radio astronomers are considering to extend FAST with several similar spherical telescopes to form a telescope array --- the FAST Array (FASTA). The initial phase of FASTA will consist of three spherical reflectors, each reflector will be similar to the existing one (we will refer to Phase-I FASTA as FASTA3 hereafter). The second phase of FASTA will add another three reflectors and bring the total number to six (similarly, we will refer to Phase-II FASTA as FASTA6 hereafter). FASTA will perform both coherent and incoherent observations of pulsars, providing a substantially higher gain than any current system, and would be able to discover a large number of potentially observable pulsars in the northern sky. In terms of sky coverage, FASTA and SKA will complement each other.

In this work, we estimated the prospect of pulsar discoveries for the FASTA in the sky area with galactic latitude $|b|<10^{\circ}$. We considered both the case  of FASTA3 and FASTA6, as well as  data analysis using the incoherent summing or coherent beam-forming methods.
Such a pulsar survey will help us understand the galactic pulsar population, especially at the low luminosity end. In section 2, we describe various pulsar population models and survey parameters used in the simulations. In section 3, we present the population synthesis results and FAST/FASTA detection prospect predicted by different models. In section 4, we summarise our conclusion and discuss future work.

\section{Method}
\label{sec:method}
\subsection{The Galactic pulsar population models}
\label{sec:dist}

There are two general types of models for the Galactic pulsar population, `snapshot' type of models and `evolutionary' type of models. Within the two types of models, there are sub-group models which apply different population distribution parameters.

A typical `snapshot' model was built up by \cite{Lorimer2006}, hereafter L06. They used the result from the Parkes Multi-beam Pulsar Survey \citep[PMPS; ][]{Manchester2001} to derive the optimal Galactic pulsar distribution in period ($P$), luminosity ($L$), Galactocentric radial distance ($R$), and Galactic scale-height ($z$). L06 treated these distributions independently, and assumed that all the canonical pulsars follow the same, fixed distribution regardless of their age.
The `snapshot' type of model is a kind of simplified model, since those distributions actually evolve with pulsar age. For example, younger pulsars, on average, are brighter and spin faster. They are also located closer to the galactic plane (associated with their progenitor stars) than older pulsars. Thus, pulsars in different age groups should have different distributions of period, luminosity, and scale height. 

The L06 type of model can be regarded as a `snapshot' of an evolving population. Despite the fact that `snapshot' type of models rely on fewer assumptions and have fewer degrees of freedom, which makes them easier to evaluate, it is still worthwhile to develop models that take evolution into account.
\cite{FK06}, hereafter FK06, presented a typical practice of constructing an `evolutionary' type of model. For comparison, FK06 additionally  provided an `unevolved' luminosity distribution which is one of the distribution model we applied in this work. Apart from FK06, a lot of efforts have also been made toward optimizing and applying evolutionary models in pulsar population synthesis, e.g. \cite{CS2006,Ridley2010,Bates2014,Rajwade2017,Huang2020}.

\subsubsection{Snapshot type of models}

The `snapshot' type of models often contains the following input parameters: (1) period distribution; (2) luminosity distribution; (3) radial density $R$ distribution; (4) scale height $z$.

The most widely accepted period distribution for `snapshot' type of canonical pulsar population models is the lognormal distribution with a mean of $\langle{\rm log}_{10}(\frac{P}{\rm{ms}})\rangle=2.7$ and a standard deviation $\sigma[{\rm log}_{10}(\frac{P}{\rm{ms}})]=0.34$, suggested by L06 based on modeling the PMPS result using \texttt{PSRPOP}\footnote{A FORTRAN package to carry out Monte Carlo simulation of the Galactic pulsar population developed by \cite{Lorimer2006} http://psrpop.sourceforge.net/}. In this work, we apply this L06 lognormal distribution for our `snapshot' type simulation of the canonical pulsar population.

For the luminosity distribution of CPs, instead of using the simple power law distribution (with a low-frequency cut-off) suggested by L06, we adopt the FK06 luminosity distribution. FK06 suggested a lognormal luminosity distribution to avoid the hard cut-off. Based on modeling the PMPS result, the mean and standard deviation of FK06 luminosity distribution are $\langle{\rm log}_{10}L\rangle=-1.1$, $\sigma[{\rm log}_{10}L]=0.9$.

For the beaming fraction of CPs in `snapshot' type of models, \cite{Emmering1989} provided a  simple approach:
\begin{equation}
f(\alpha, \rho)=2\left(\frac{1}{4 \pi} \int_{\theta_l}^{\theta_u} 2 \pi \sin \theta d \theta\right)=\cos \theta_l-\cos \theta_u ,
\end{equation}
where $\alpha$ is the inclination angle, $\rho$ is the beam radius, $\theta_l={\rm max}(0, \alpha-\rho)$ and $\theta_u={\rm min}(\pi/2, \alpha+\rho)$.
One empirical way to estimate the beam radius $\rho$ provided by \cite{Gould1994}, is to treat $\rho$ as a function of the pulsar rotating period,
$\rho(P) = 5.4^{\circ} (\frac{P}{\rm s})^{-1/2}$, while $\alpha$ is a uniformly chosen random value between (0$^{\circ}$, 90$^{\circ}$). We will further discuss the effects of different beaming fraction models in Section 2.2 where we describe the role of calibration surveys for our simulation.

For the radial density distribution of CPs, L06 suggested a gamma distribution (applying NE2001 electron density model):
\begin{equation}\label{eq:rdist}
  f(R) = A \left(\frac{R}{R_{\odot}}\right)^{B}
  \exp \left(-C\left[\frac{R-R_{\odot}}{R_{\odot}}\right]\right),
\end{equation}
where $A=41$, $B=1.9$, and $C=5.0$. This relation implies that the pulsar density is zero at $R=0$, which is inconsistent with our understanding of the galaxy structure. To avoid this discrepancy and obtain nonzero density at $R=0$, \cite{YK2004}, hereafter YK04, included an additional parameter $R_1$ and used a shifted Gamma function, replacing $R$ and $R_\odot$ in Eq.\,(\ref{eq:rdist}) by $X=R+R_1$ and $X_\odot =R_\odot +R_1$. With this model, YK04 obtained the best-fit result with a parameter set of $ A=37.6$, $B=1.64$,  $C=4.01$ and $R_1=0.55$\,kpc. In this work, we compare FAST and FASTA simulated pulsar survey detection prospects using both L06 and YK04 radial distributions.

For the scale height ($z$) distribution of CPs, L06 found an optimal scale height of 180\,pc using NE2001 electron density model. This number is significantly lower than the expected value of 300$-$350\,pc from independent studies of the local CPs population \citep[e.g. ][]{Mdzinarishvili2004}. When applying the `smooth' electron density distribution model \citep{Lyne1985}, the optimal scale height, 330\,pc, is consistent with the observed local CPs population \citep{Lorimer2006}; this is the value we used in this work.


To describe the Galactic MSP population with the `snapshot' type of models, the approaches are similar to the aforementioned CPs cases. Note that the modeled parameters for MSPs are limited by the relatively small sample size of the currently known MSP population.
For the period distribution of MSPs, \cite{Lorimer2015}, hereafter L15, suggested a lognormal distribution with a mean $\langle {\rm log}_{10} (\frac{P}{\rm{ms}}) \rangle = 0.65$ and a standard deviation $\sigma[{\rm log}_{10}(\frac{P}{\rm{ms}})]=0.25$. For the luminosity distribution of MSPs, following \cite{Smits2009b}, we adopt the same luminosity distribution of canonical pulsars. The beaming fraction of MSPs we used is $f(\alpha, \rho)=\cos \theta_l-\cos \theta_u$, where $\theta_l={\rm max}(0, \alpha-\rho)$ and $\theta_u={\rm min}(\pi/2, \alpha+\rho)$. As suggested by \cite{Kramer1998}, the beam radius, $\rho$, is a constant value for MSPs. Here we apply $\rho=31^{\circ}$ which corresponds to $\rho(P) = 5.4^{\circ} (\frac{P}{\rm s})^{-1/2}, P=30$\,ms . We use 500\,pc as the scale height for the Galactic MSP population as suggested by \cite{Smits2009b}. For $R$ distribution, we apply both the L06 and the YK04 radial distribution, respectively.

The input model parameters we applied to model the Galactic CP and MSP populations using the `snapshot' type of models are summarised in Table~\ref{tab:Snappara}.

\subsubsection{Evolutionary type of models}

The `evolutionary' type of models also  use the following parameters to describe the Galactic pulsar population:
(1) period distribution; (2) luminosity distribution; (3) radial density $R$ distribution; (4) scale height $z$.
The main difference between the evolutionary type of models and the `snapshot' type of models is that for the evolutionary type of models, those pulsar parameters are all described as a function of the pulsar age. In this work, we apply 1\,Gyr for the maximum age of the canonical pulsar population, and 5\,Gyr for the maximum age of the MSP population based on the current distribution of observational pulsar characteristic age \citep{Rajwade2017}.

The period distribution for the evolutionary type of models is described by: (1) the initial (birth) spin period distribution; (2) the pulsar spin-down model; (3) the pulsar age.
FK06 developed a typical `evolutionary' model, and found the optimal initial period distribution for canonical pulsars follows a normal distribution with the mean $\langle P_0 \rangle = 300$\,ms. The observed pulsar rotating period at the present day can be evaluated using the initial period and the age of the pulsar.
The spin-down model depends on the braking index $n$ and the magnetic field strength $B$.
In this work, we adopt the FK06 spin-down model and initial period distribution for canonical pulsars. We note that other `evolutionary' type models may employ a different spin-down model and initial period distribution \citep[e.g.][]{CS2006}.

The luminosity distribution is also age-dependent. Since the radiated energy of pulsars is thought to be  originated from the loss of rotational energy, the luminosity distribution can be modeled as a function of its period ($P$, in units of s), and its period derivative ($\dot{P}$, in units of 10$^{-15}$\,s\,s$^{-1}$). It is commonly assumed to be a power-law: $L = \gamma P^{a} \dot{P}^{b}$, while the values of $a$, $b$, $\gamma$ may vary among different models (e.g. \citealt{Lyne1975,Vivekanand1981,FK06,Rajwade2017}). The original `evolutionary' model, FK06, found the optimal value for $a$, $b$, and $\gamma$ are: $a=-1.5$, $b=0.5$, and $\gamma=0.18$
based on the PMPS pulsar sample.
In this work, we follow the FK06 model.
As a comparison, we also test the luminosity distribution derived from the fan-beam model \citep{Wang2014,Huang2020}.
The luminosity function of the fan-beam model not only depends on $P$ and $\dot{P}$, but also the emission geometry in the term of $L = \kappa \frac{W}{P} P^{q-4} \dot{P} \rho_{\rm peak}^{2q-6}$, where $\frac{W}{P}$ is the duty cycle and $\rho_{\rm peak}$ is the radial distance between the magnetic pole and the emission direction accounting for the pulse peak in units of degree. Following \cite{Huang2020}, we adopt $\kappa = 10^{2.75}$, $\frac{W}{P}=5$\%, $q=1.25$ in our simulation. Unlike the case of the conal beam, in the fan beam model, the impact angle between the line of sight (LOS) and the magnetic axis  may extend to 90$^{\circ}$, which means our LOS would sweep across at least one emission beam from either one pole or the other pole, so the beaming fraction is always 1.
While the default evolutionary model applied an empirical beaming fraction given by \cite{TM1998}, $f(P) = 0.09({\rm log}{P} -1)^{2} + 0.03$, which is a function of the pulsar period.

The current position of each simulated pulsar in our Galaxy is determined by its initial birth position (following the $z$ and $R$ birth distribution), and its birth velocity. According to FK06, the initial birth $z$ distribution follows an exponential function with a relatively small scale height of 50\,pc. This is consistent with the distribution of supernovae where massive stars end their lives. While for the $R$ distribution, FK06 applied the YK04 radial distribution as the birth radial distribution. In this work, we have tried the L06 and the YK04 radial distributions as the birth radial distribution. For the pulsar birth velocity distribution, we assume a Gaussian distribution centred on 0\,km/s with a width of 265\,km/s for the birth velocity for each of the x, y and z directions following \cite{Rajwade2017}. The pulsar position will then evolve from its initial position according to its velocity and age, according to the model of the Galaxy gravitational potential \citep{Carlberg1987, Kuijken1989}.


For evolutionary MSP population models, we  adopted the L15 distribution for their initial birth period, which is a lognormal distribution with average $\langle {\rm ln(\frac{P}{\rm{ms}})} \rangle = 1.5$ and a standard deviation $\sigma[{\rm ln}(\frac{P}{\rm{ms}})]=0.58$. We have tried two types of $L$ distribution: (1) the FK06 fixed lognormal distribution with $\langle{\rm log}_{10}L\rangle=-1.1$, $\sigma[{\rm log}_{10}L]=0.9$; (2) the FK06 evolved power-law distribution model, $L = \gamma P^{a} \dot{P}^{b}$, where $a=-1.4$, $b=0.5$, and $\gamma=0.009$ \citep{Rajwade2017}.
The initial birth $z$ distribution we adopted is the same as the evolutionary model for canonical pulsars (CPs), which is an exponential function with a scale height of 50\,pc (FK06). For $R$ distribution, similar with our CPs simulation, we have tested both the L06 and the YK04 radial distributions as the birth radial distribution.
For the pulsar birth velocity distribution, following \cite{Rajwade2017}, we assume a Gaussian distribution centred on 0\,km/s with a width of 80\,km/s for the birth velocity for each of the x, y and z directions.

The input model parameters we applied to model the Galactic CP and MSP populations using the evolutionary type of models are summarised in Table~\ref{tab:Evopara}.

\begin{table}[htbp]
\centering
\caption{The model parameters used in snapshot mode simulation of CPs and MSPs.}
\label{tab:Snappara}
\begin{tabular}{lcc}
\hline
Parameter & CP & MSP  \\
\hline
Spin period distribution & L06 lognormal & L15 lognormal\\
$\langle{\rm log}_{10}(\frac{P}{\rm{ms}})\rangle$ & 2.7 & 0.65 \\ 
$\sigma[{\rm log}_{10}(\frac{P}{\rm{ms}})]$ & 0.34 & 0.25 \\ 
\\
Luminosity distribution & FK06 lognormal & FK06 lognormal\\
$\langle{\rm log}_{10}L\rangle$  & -1.1 & -1.1 \\
$\sigma[{\rm log}_{10}L]$ & 0.9 & 0.9 \\
\\
Galactic $z$-scale height & 330\,pc & 500\,pc \\
Radial distribution Model A & L06 & L06 \\
Radial distribution Model B & YK04 & YK04 \\
\\
Spectral index Distribution & Gaussian & Gaussian \\
$\langle a \rangle$ & -1.4 & -1.4 \\
$\sigma_{a}$ & 0.9 & 0.9 \\

\\
Population scaling surveys & PMPS+SWIL+SWHL & PMPS+SWIL+SWHL \\
Detected pulsar number in scaling surveys & 1214 & 48\\
\hline
\end{tabular}
\end{table}

\begin{table}[htbp]
\centering
\caption{The model parameters used in evolutionary mode simulation of CPs and MSPs.}
\label{tab:Evopara}
\begin{tabular}{lccc}
\hline
Parameter & CP & MSP  \\
\hline
Initial spin period distribution & FK06 Gaussian & L15 lognormal\\
$\langle \frac{P}{\rm{ms}} \rangle$  & 300  & --- \\ 
$\sigma[\frac{P}{\rm{ms}}]$ &150 & --- \\ 
$\langle {\rm ln}\frac{P}{\rm{ms}} \rangle$  &--- & 1.5 \\
$\sigma[{\rm ln}(\frac{P}{\rm{ms}})]$ &--- & 0.58 \\
\\
Luminosity distribution model A & FK06 evolutionary & FK06 evolutionary\\
 & $L=0.18 P^{-1.5} \dot{P}^{0.5}$ & $L=0.009 P^{-1.4} \dot{P}^{0.5}$ \\
Luminosity distribution model B & Fan beam model & FK06 lognormal\\
\\
Initial Galactic $z$-scale height & 50\,pc & 50\,pc \\
Radial distribution Model A & L06 & L06 \\
Radial distribution Model B & YK04 & YK04 \\
1-D velocity dispersion  &  265~km~s$^{-1}$    &   80~km~s$^{-1}$   \\
Maximum initial age & 1~Gyr & 5~Gyr \\
\\
Spectral index Distribution & Gaussian & Gaussian \\
$\langle a \rangle$ & -1.4 & -1.4 \\
$\sigma_{a}$ & 0.9 & 0.9 \\
\\
Pulsar spin-down model & FK06 & FK06 \\
Beam alignment model & orthogonal & orthogonal \\
Braking Index & 3.0 & 3.0 \\
Initial B-field distribution & Log-normal & Log-normal \\
$\langle \log_{10}{\rm B(G)} \rangle$ &12.65 & 8.0 \\
${\rm std}(\log_{10}{\rm B(G)})$ & 0.55 & 0.55\\
\\
Population scaling surveys & PMPS+SWIL+SWHL & PMPS+SWIL+SWHL \\
Detected pulsar number in scaling surveys & 1214 & 48\\
\hline
\end{tabular}
\end{table}

\subsection{FAST and FASTA pulsar survey simulations}
In this work, we used \texttt{PsrPopPy} to perform the simulation. The simulation process of \texttt{PsrPopPy} contains two main steps: (1). generate synthetic pulsar populations of the Milky Way Galaxy; (2). perform simulated surveys on the synthetic pulsar population using corresponding survey parameters.

When \texttt{PsrPopPy} generates synthetic pulsar populations of the Milky Way Galaxy, the overall idea of how it controls the total number of the pulsar populations is reversed to how we think about this question intuitively. It does not first assume a total Galactic neutron star population with a beaming fraction to get the population of the pulsars that beaming to us, and then estimate how many pulsars we can detect. 
On the contrary, it uses the real numbers of discoveries from existing pulsar surveys and the corresponding survey parameters to calibrate the total number of the synthesized pulsar population.
The calibration surveys need to have a good enough completeness. A commonly used calibration pulsar survey is the PMPS survey which discovered over a thousand pulsars in a uniform setup.
For example, \cite{Rajwade2017} applied the same detection threshold and observing ranges of the PMPS to their synthesized pulsar population with a pulsar detection sample size of 1065. In this case, \texttt{PsrPopPy} keeps generating new simulated pulsars until the generated population contains exactly 1065 pulsars detectable by the same setup as PMPS. All the pulsars generated in this process, if they are beaming toward us, are stored as the whole synthetic pulsar population for further analysis, no matter whether they are detectable by PMPS or not. While the simulated pulsars which are not beaming toward us are all left.
In such a way, the total population mostly depends on the calibration surveys, while the beaming fraction is not a key dominant factor that directly affects the simulated detection number of pulsars. It affects the simulation result in a relatively secondary way due to its relation with other characteristics of pulsars, namely the pulsar rotating period $P$. Since longer period pulsars have smaller beaming fraction, the effect is like shifting the period distribution toward the shorter $P$ end.

\texttt{PsrPopPy} can adopt multiple calibration surveys together as a whole when generating synthetic pulsar population. Here, following \cite{Huang2020}, we adopt a combination of PMPS, together with two Swinburne pulsar surveys (SWIL, \citealt{Edwards2001}; and SWHL, \citealt{Jacoby2007}) as calibration surveys. This provides us a sample of 1214 canonical pulsars \citep{Huang2020} and 48 MSPs \citep{Lorimer2015}.

In the second step of the simulation, after generating a realisation of the Galactic pulsar population (for canonical pulsars or MSPs, respectively), we then simulate pulsar surveys using FAST and FASTA instrument parameters.
The FAST survey parameters (including gain, system temperature, bandwidth, centre frequency, time and frequency resolution, etc.) are set according to FAST performance and survey description papers \citep[e.g. ][]{Li2018,Jiang2019,Jiang2020,Han2021}. For FASTA survey, we apply the same survey parameter as FAST Galactic plane survey \citep{Han2021}.
In the ideal coherent beam-forming process, an array consisting of $N$ antennae would have a sensitivity $N$ times better ($G_{\rm array} = N \times G_{\rm antenna} $). In practice, we considered a phasing efficiency parameter of 0.94 \citep{Chen2021} when estimating the gain value ($G_{\rm array} = 0.94 \times N \times G_{\rm antenna}$).
The gain value for FAST is 16\,K/Jy, thus, the estimated gain for FASTA3 and FASTA6 are $\sim 45$\,K/Jy and $\sim 90$\,K/Jy, respectively.

For the cases of incoherent sum, the gain of an array consisting of $N$ antennae would be $G_{\rm array} = \sqrt{N} \times G_{\rm antenna} $. Therefore, the incoherent gain for FASTA3 and FASTA6 are $\sim 28$\,K/Jy and $\sim 40$\,K/Jy.

We note that the sensitivity of FAST drops drastically when the zenith angle (ZA) is greater than 26.4$^{\circ}$. According to \cite{Jiang2019}, the gain of FAST is 16\,K/Jy when ZA$<26.4^{\circ}$, while it decreases as a linear function of ZA and becomes 11\,K/Jy when ZA$=40^{\circ}$. We have taken this effect into account in our simulation for both FAST and FASTA.

All the survey parameters we adopted for FAST and FASTA are summarised in Table~\ref{tab:survP}. We use FASTA3 and FASTA6 to represent the coherent beam-forming cases. While for incoherent summing cases, we use FASTA3-i and FASTA6-i, respectively.

\begin{table}[htbp]
  \centering
  \caption{System parameters used in the simulated survey with FAST and FASTA}
  \begin{ThreePartTable}
    \begin{tabular}{lccccc}
     \hline
      & FAST &  FASTA3-i  &  FASTA6-i  &  FASTA3  &  FASTA6  \\
     \hline
    Survey degradation factor             & 0.8   & 0.8   & 0.8   & 0.8   & 0.8 \\
    Gain (K/Jy)\tnote{\textdagger}        & 16    & 28    & 40    & 45    & 90 \\
    Integration time (s)                  & 300   & 300   & 300   & 300   & 300 \\
    Sampling time (ms)                    & 0.2   & 0.2   & 0.2   & 0.2   & 0.2 \\
    System temperature (K)                & 20    & 20    & 20    & 20    & 20 \\
    Centre frequency (MHz)                & 1250  & 1250  & 1250  & 1250  & 1250 \\
    Bandwidth (MHz)                       & 400   & 400   & 400   & 400   & 400 \\
    Channel bandwidth (MHz)               & 0.125 & 0.125 & 0.125 & 0.125 & 0.125 \\
    Number of polarizations               & 2     & 2     & 2     & 2     & 2 \\
    Full-width half maximum (arcmin)      & 3     & 3     & 3     & 3     & 3 \\
    Minimum RA (deg)                      & 0     & 0     & 0     & 0     & 0 \\
    Maximum RA (deg)                      & 360   & 360   & 360   & 360   & 360 \\
    Minimum DEC (deg)                     & -14   & -14   & -14   & -14   & -14 \\
    Maximum DEC (deg)                     & 65    & 65    & 65    & 65    & 65 \\
    Minimum abs(Galactic latitude) (deg)  & 0     & 0     & 0     & 0     & 0 \\
    Maximum abs(Galactic latitude) (deg)  & 10    & 10    & 10    & 10    & 10 \\
    Fractional sky coverage (0-1)         & 1     & 1     & 1     & 1     & 1 \\
    Minimum signal-to-noise               & 9     & 9     & 9     & 9     & 9 \\

    \hline
    \end{tabular}%
  \label{tab:survP}
\begin{tablenotes}
    \item[\textdagger] The values listed in this row are the gain value when pointing to ZA$<26.4^{\circ}$. For $26.4^{\circ}<$ZA$<40^{\circ}$, the gain value we applied is a linear function of ZA, which is $G=G_{\rm zenith} \times\left[1-0.0216({\rm ZA} -26.4^{\circ})\right]$.
\end{tablenotes}
\end{ThreePartTable}
\end{table}%

\section{Result}

\subsection{Detection prospects of Galactic CPs/MSPs with FASTA}
\label{sec:prospectMW}
With each set of the synthetic pulsar population model parameters summarised in Table~\ref{tab:Snappara} and Table~\ref{tab:Evopara}, we simulate 20 realisations of pulsar populations.
While for each realisation, we perform 20 times survey simulations for each set of survey parameters listed in Table~\ref{tab:survP}. Therefore, for each parameter combination, we run 400 times simulations in total to get the detection prospects for either canonical pulsars or MSPs. The results are summarised in Table~\ref{tab:resultCPs} and Table~\ref{tab:resultMSPs}.

For canonical pulsars that FAST is able to detect, the prediction numbers given by different models range from $\sim$ 4200 to 8460, while for the most sensitive FASTA6 coherent beamform mode, the predicted detectable CPs range from $\sim$ 8380 to 17330. These prediction numbers all include currently known pulsars. Therefore, when we consider the number of new CPs discoveries, we need to subtract $\sim$ 1000 known Galactic-plane CPs in the sky field of FAST/FASTA survey.
For MSPs, FAST detection prospect given by different models range from $\sim$ 520 to 1480, and the number of FASTA6 detectable MSPs range from $\sim$ 1140 to 4680. Similar to the case of CPs, these prediction numbers include known MSPs, and when considering new MSPs discoveries, $\sim$ 100 known MSPs in the survey field need to be subtracted.

\begin{figure*}
   \centering
   \includegraphics[scale=0.6, angle=0]{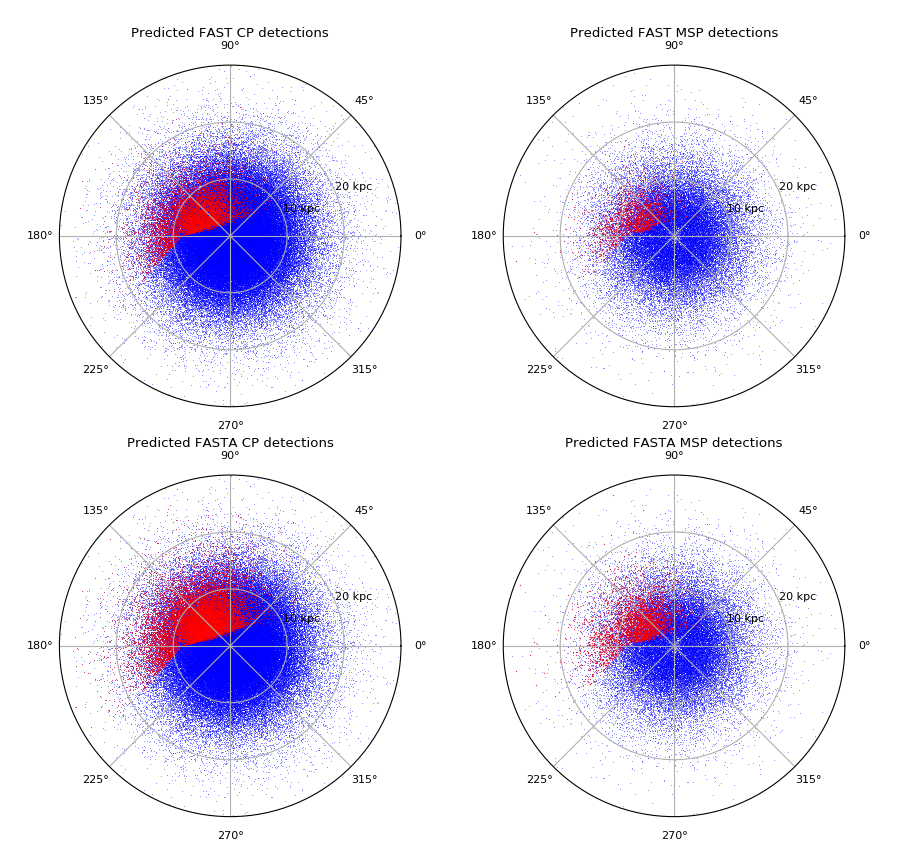}
   \caption{Galactic distribution of the simulated pulsar population (blue dots) and predicted FAST/FASTA6 survey detection (red dots) in polar coordinates (top-view from the Galactic North pole). The synthetic pulsar population of this example plot is generated based on snapshot mode with YK04 radial density distribution.} 
   \label{Fig:Galpol}
\end{figure*}

Figure~\ref{Fig:Galpol} shows the Galactic distribution of the simulated pulsar population and predicted survey detection in polar coordinates. We choose the predicted result from the current FAST and the most sensitive FASTA6 coherent beamform mode to illustrate. The synthetic pulsar population of this example plot is generated based on snapshot mode with YK04 radial density distribution.

\begin{table}[htbp]
\centering
\caption{Model parameters and the results for FAST and FASTA simulated survey on Galactic CPs.}
\label{tab:resultCPs}
\begin{ThreePartTable}
\begin{tabular}{llllll}
    \hline
     Simulation mode & P-dist & L-dist &  R-dist &  \multicolumn{2}{c}{Survey predicted detections~(1$\sigma$)\tnote{\textdagger}}  \\
    \hline
    \multirow{10}{*}{Snapshot} &
    \multirow{10}{*}{L06 lognormal} &
    \multirow{10}{*}{\makecell{FK06 lognormal}}&
    \multirow{5}{*}{L06}  & FAST & 6280$\pm$200      \\
      &  &  &  & FASTA3-i & 8200$\pm$250  \\
      &  &  &  & FASTA6-i & 9620$\pm$290  \\
      &  &  &  & FASTA3   & 10080$\pm$310  \\
      &  &  &  & FASTA6   & 12940$\pm$400  \\
      \cline{4-6}
      &  &  &  \multirow{5}{*}{YK04} & FAST & 8460$\pm$230      \\
      &  &  &  & FASTA3-i & 11010$\pm$290  \\
      &  &  &  & FASTA6-i & 12810$\pm$340  \\
      &  &  &  & FASTA3 & 13430$\pm$360  \\
      &  &  &  & FASTA6 & 17330$\pm$460  \\
    \hline
    \multirow{20}{*}{Evolutionary} &
    \multirow{20}{*}{\makecell{FK06 lognormal}} &
    \multirow{10}{*}{\makecell{ $L=0.18 P^{-1.5} \dot{P}^{0.5}$  \\ (B=12.65, FK06)}} &
    \multirow{5}{*}{L06}  & FAST & 4200$\pm$140      \\
      &  &  &  & FASTA3-i & 5400$\pm$170  \\
      &  &  &  & FASTA6-i & 6230$\pm$200  \\
      &  &  &  & FASTA3 & 6520$\pm$210  \\
      &  &  &  & FASTA6 & 8380$\pm$270  \\
      \cline{4-6}
      &  &  &  \multirow{5}{*}{YK04} & FAST & 5080$\pm$180      \\
      &  &  &  & FASTA3-i & 6520$\pm$240  \\
      &  &  &  & FASTA6-i & 7560$\pm$280  \\
      &  &  &  & FASTA3   & 7910$\pm$290  \\
      &  &  &  & FASTA6   & 10160$\pm$370  \\
    \cline{3-6}
    & &
    \multirow{10}{*}{\makecell{ Fan beam model  \\ \citep{Huang2020}}} &
    \multirow{5}{*}{L06}  & FAST & 4460$\pm$170      \\
      &  &  &  & FASTA3-i & 6010$\pm$240  \\
      &  &  &  & FASTA6-i & 7230$\pm$280  \\
      &  &  &  & FASTA3 & 7670$\pm$290  \\
      &  &  &  & FASTA6 & 10640$\pm$390  \\
      \cline{4-6}
      &  &  &  \multirow{5}{*}{YK04} & FAST & 5650$\pm$220  \\
      &  &  &  & FASTA3-i & 7600$\pm$270  \\
      &  &  &  & FASTA6-i & 9070$\pm$320  \\
      &  &  &  & FASTA6 & 9590$\pm$340  \\
      &  &  &  & FASTA6 & 13290$\pm$470  \\
  \hline
\end{tabular}
\begin{tablenotes}
    \item[\textdagger] The predicted detection number with 1$\sigma$ distribution dispersion.
\end{tablenotes}
\end{ThreePartTable}
\end{table}

\begin{table}[htbp]
\centering
\caption{Model parameters and the results for FAST and FASTA simulated survey on Galactic MSPs.}
\label{tab:resultMSPs}
\begin{ThreePartTable}
\begin{tabular}{llllll}
   \hline
     Simulation mode & P-dist & L-dist &  R-dist & \multicolumn{2}{c}{Survey predicted detections~(1$\sigma$)\tnote{\textdagger}} \\
    \hline
    \multirow{10}{*}{Snapshot} &
    \multirow{10}{*}{L15 lognormal} &
    \multirow{10}{*}{\makecell{FK06 lognormal}}&
    \multirow{5}{*}{L06}  &  FAST & 620$\pm$100     \\
      &  &  &  & FASTA3-i & 800$\pm$130  \\
      &  &  &  & FASTA6-i & 940$\pm$150  \\
      &  &  &  & FASTA3   & 990$\pm$160  \\
      &  &  &  & FASTA6   & 1300$\pm$200  \\
      \cline{4-6}
      &  &  &  \multirow{5}{*}{YK04} & FAST & 820$\pm$90 \\
      &  &  &  & FASTA3-i & 1090$\pm$120  \\
      &  &  &  & FASTA6-i & 1270$\pm$140  \\
      &  &  &  & FASTA6 & 1340$\pm$150  \\
      &  &  &  & FASTA6 & 1750$\pm$200  \\
    \hline
    \multirow{20}{*}{Evolutionary} &
    \multirow{20}{*}{\makecell{L15 model as initial \\ $P$ distribution}} &
    \multirow{10}{*}{\makecell{FK06 lognormal}}&
    \multirow{5}{*}{L06}  & FAST & 520$\pm$60     \\
      &  &  &  & FASTA3-i & 700$\pm$80 \\
      &  &  &  & FASTA6-i & 820$\pm$100 \\
      &  &  &  & FASTA3 & 860$\pm$100 \\
      &  &  &  & FASTA6 & 1140$\pm$140 \\
      \cline{4-6}
      &  &  &  \multirow{5}{*}{YK04} & FAST & 560$\pm$80 \\
      &  &  &  & FASTA3-i & 740$\pm$110 \\
      &  &  &  & FASTA6-i & 880$\pm$130 \\
      &  &  &  & FASTA6 & 930$\pm$140 \\
      &  &  &  & FASTA6 & 1240$\pm$190 \\
    \cline{3-6}
    & &
    \multirow{10}{*}{\makecell{ $L=0.009 P^{-1.4} \dot{P}^{0.5}$  \\ (\citealt{Rajwade2017})}} &
    \multirow{5}{*}{L06}  & FAST & 1290$\pm$230     \\
      &  &  &  & FASTA3-i & 1960$\pm$350 \\
      &  &  &  & FASTA6-i & 2530$\pm$450 \\
      &  &  &  & FASTA6 & 2740$\pm$490 \\
      &  &  &  & FASTA6 & 4210$\pm$760 \\
      \cline{4-6}
      &  &  &  \multirow{5}{*}{YK04} & FAST & 1480$\pm$190 \\
      &  &  &  & FASTA3-i & 2230$\pm$280 \\
      &  &  &  & FASTA6-i & 2830$\pm$360 \\
      &  &  &  & FASTA6 & 3070$\pm$390 \\
      &  &  &  & FASTA6 & 4680$\pm$610 \\
  \hline
\end{tabular}
\begin{tablenotes}
    \item[\textdagger] The predicted detection number with 1$\sigma$ distribution dispersion.
\end{tablenotes}
\end{ThreePartTable}
\end{table}

\subsection{Detection prospects of radio pulsars in M31 with FASTA}
\label{sec:prospectM31}
Up to now, astronomers haven't discovered radio pulsars in the nearby galaxy M31. FASTA will have great potential to make the breakthrough discovery in the detection of the first pulsar in M31.
According to \cite{Savino2022}, the distance of M31 is 776$\pm$22\,kpc.
Considering a time integration of 2 hours for each pointing (corresponding to a total survey time of $\sim$37 hours with 10-min overhead, covering M31 in 17 pointings), and a bandwidth of 400\,MHz, dual-polarisation observation, FAST will be able to detect pulsars with L-band luminosity greater than $670$\,mJy\,kpc$^2$ at the distance of M31. While for FASTA3-i, FASTA6-i, FAST3, and FASTA6, the corresponding luminosity limit at L-band would be 380\,mJy\,kpc$^2$, 270\,mJy\,kpc$^2$, 240\,mJy\,kpc$^2$, and 120\,mJy\,kpc$^2$, respectively.
We can then estimate the detection prospect of radio pulsars in M31 based on how many pulsars in M31 have luminosity above the detection limit of each configuration. We assume the M31 has a similar pulsar population to our Galaxy.

We considered three types of population models: 1) the snapshot model with FK06 lognormal luminosity distribution; 2) the evolutionary model with FK06 luminosity distribution; 3) the evolutionary model with fanbeam luminosity distribution. These models are the same as what we used in the previous sections when we estimated the detection prospect of Galactic canonical pulsars, but we did not distinguish the L06 and YK04 radial distribution for M31 pulsar detection prospect estimation.

For the snapshot model with FK06 lognormal luminosity distribution, the corresponding pulsar detection numbers of FAST, FASTA3-i, FASTA6-i, FAST3, FASTA6 are 1, 3, 7, 9, 28, respectively. For the evolutionary model with FK06 luminosity distribution, the corresponding pulsar detection numbers are 12, 29, 47, 60, 164, respectively.
For the evolutionary model with fanbeam luminosity distribution, the corresponding pulsar detection numbers are 34, 55, 77, 92, 188, respectively.

Although the detection prospect predicted by those three models are quite different,  they all show the same trend as the instrument sensitivity increases, especially in the case of FASTA6. Thus, there is a good chance to detect a significant amount of radio pulsars in M31.

\subsection{Survey time estimation}
The total observable sky for FAST with galactic latitude $|b|<10^{\circ}$ is 4035\,deg$^2$. If FASTA makes use of the Phased Array Feed (PAF) with FWHM of 30', then the survey will need 20550 pointings in total. As the planned integration time is 5 minutes, the entire survey will need to take 1713 observing hours.

\begin{figure*}
   \centering
   \includegraphics[scale=0.85, angle=0]{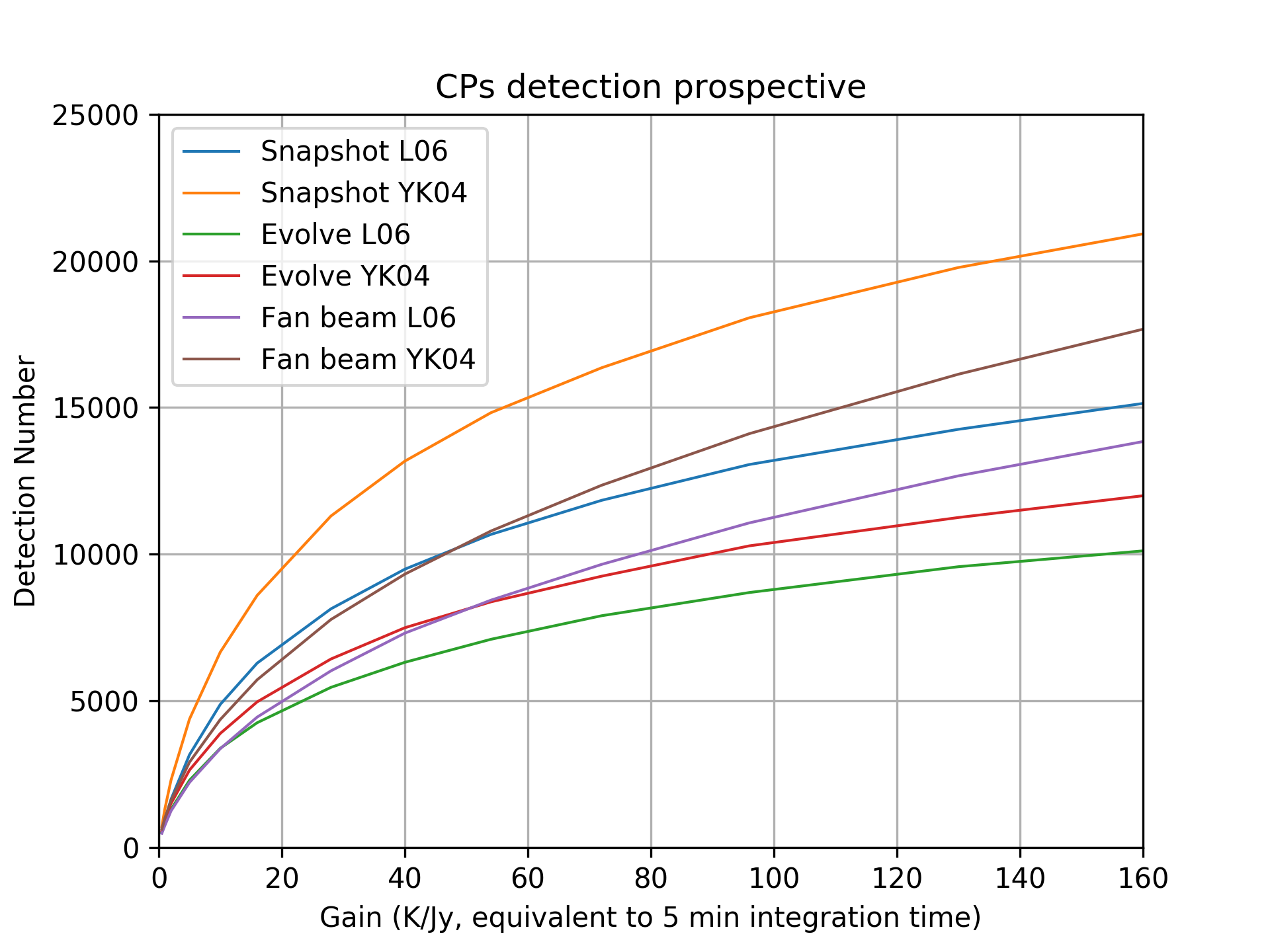}
   \caption{Number of canonical pulsars detected in simulated surveys with 5-min integration time for different gain values, remaining parameters we used in these simulations are all from Table~\ref{tab:survP}. For other integration times that we want to compare (e.g. 10, 15, 20 min as listed in the top panel of Table~\ref{tab:NumvsGain}), we can calculate the corresponding gain value that equivalent to 5-min integration time, $G_{\rm equiv} = G \times \sqrt {t_{\rm int} / 5\rm{min}}$. Different color lines indicate different pulsar population models.}
   \label{Fig:NumvsGain}
\end{figure*}

If we include 10 minutes overhead for each pointing, the survey will take 5138 hours in total. Since the overhead will take a significant amount of time for each pointing, it's natural for us to raise a question --- what if we increase the observing integration time for each pointing, will it improve the return-cost-rate of the survey in terms of the pulsar detection number vs. total survey time?
To explore the answers, we run simulations by setting the observing integration time to 10, 15, 20 minutes along with different combinations of pulsar population model and survey configuration (i.e. different gain) listed in Section\,\ref{sec:prospectMW}. According to the radiometer equation, the telescope sensitivity is proportional to the instrument gain and the square-root of the integration time. In order to make it more concise, we convert different integration time to a gain-increment equivalent to 5-min integration time, $G_{\rm equiv} = G \times \sqrt {t_{\rm int} / 5\rm{min}}$. The corresponding equivalent gain values are listed in the top panel of Table~\ref{tab:NumvsGain}. We then simulate CPs detection numbers as a function of different gain values (ranging from 0.5 to 160) for different population models, with 5-min integration time. The result can be found in Figure~\ref{Fig:NumvsGain}.

Using the snapshot model with L06 radial distribution (blue line in Figure~\ref{Fig:NumvsGain}) as  an example, we list the simulated pulsar detection numbers with different observing integration time and different survey configurations in the middle panel of Table~\ref{tab:NumvsGain}. We also convert those numbers to percentages relative to FAST 5-min integration result (the bottom panel of Table~\ref{tab:NumvsGain}), so that we can get a more straightforward idea of the detection number increment. Comparing the survey time increment percentage and detection number increment percentage, we find that 5-min integration time is still the most efficient survey setting.

\begin{table}[htbp]
  \centering
  \caption{Simulated pulsar detection numbers using the snapshot model with L06 radial distribution.}
    \begin{tabular}{lcccccc}
    \hline
          & Total survey time & \multicolumn{5}{c}{Gain (equivalent to 5-min integration time)} \\
          & (hours) & FAST & FAST3-i & FAST6-i & FAST3 & FAST6 \\
    \hline
    ~5 min & 5138  & 16    & 28    & 40    & 45    & 90  \\
    10 min & 6850  & 23    & 40    & 57    & 64    & 127  \\
    15 min & 8563  & 28    & 48    & 69    & 78    & 156  \\
    20 min & 10275 & 32    & 56    & 80    & 90    & 180  \\
    \hline
          & Total survey time & \multicolumn{5}{c}{Detection numbers} \\
          & (hours) & FAST & FAST3-i & FAST6-i & FAST3 & FAST6 \\
    \hline
    ~5 min & 5138  & 6287  & 8160  & 9479  & 9943  & 12758 \\
    10 min & 6850  & 7420  & 9437  & 10855 & 11336 & 14169 \\
    15 min & 8563  & 8129  & 10240 & 11688 & 12146 & 15026 \\
    20 min & 10275 & 8654  & 10806 & 12278 & 12755 & 15623 \\
    \hline
          & Total survey time  & \multicolumn{5}{c}{Detection numbers (\% relative to FAST 5\,min)} \\
          & (hours) & FAST & FAST3-i & FAST6-i & FAST3 & FAST6 \\
    \hline
    ~5 min & 100\% & 100\% & 130\% & 151\% & 158\% & 203\% \\
    10 min & 133\% & 118\% & 150\% & 173\% & 180\% & 225\% \\
    15 min & 167\% & 129\% & 163\% & 186\% & 193\% & 239\% \\
    20 min & 200\% & 138\% & 172\% & 195\% & 203\% & 248\% \\
    \hline
    \end{tabular}%
  \label{tab:NumvsGain}%
\end{table}%

\section{Discussion and conclusion}
\label{sec:discussion}

In this work, we estimated pulsar detection prospects for simulated FAST and FASTA pulsar surveys with galactic latitude $|b|<10^{\circ}$, using various pulsar population models.
We tested models from both the snapshot and evolutionary types, with different combinations of distribution parameters.
Our results indicate that FASTA could detect around ten thousand canonical pulsars and well-over thousands of millisecond pulsars.
Additionally, we estimated the yield of searching for pulsars in the nearby spiral galaxy M31 using FASTA, and found that it has a potential to discover around a hundred new radio pulsars.
Furthermore, we also found that the most efficient observational settings in terms of the observing time and discovery number for a Galactic-plane pulsar survey is to apply 5-min integration time with a PAF.

We estimated the pulsar detection prospect of using both the incoherent sum or coherent beam-forming data analysis method for FASTA data. The baseline design of FASTA has not been determined yet, but as a first-order approximation, we can consider the maximum baseline of FASTA to be $\sim$300\,km. Since the baseline is 1000 times longer than the effective diameter of FAST (300\,m), we need to form $10^{6}$ coherent beams to cover the area of one current FAST beam. Therefore, to cover the sky area of the proposed survey, the number of beams we need to form is on the order of $10^{12}$. This will inevitably result in a significant increase in data processing demand, which seems to be infeasible with current computational instruments and techniques. Nevertheless, we can remain hopeful that the requirement can be met in a few decades since Moore's Law suggests that the computational capability would increase one order of magnitude in 5 to 7 years, and there might be other potential technique revolution in the coming future. While for the beginning, a pulsar survey with incoherently summing data would be more realistic, and we can first apply coherent beam-forming data for some specific sky areas such as M31 or selected globular clusters.

With the high sensitivity of FASTA, it is likely that more pulsars (especially long-period pulsars, intermittent pulsars and RRATs) could be detected when adopting additional single pulse search.
According to existing pulsar surveys with large spherical radio telescopes, \cite{Deneva2009} found that the single pulse search led to 13\% extra pulsar discoveries from Arecibo PALFA survey, \cite{Han2021} found a 14\% extra yield from FAST GPPS survey, while as a drift scan survey with each point source drifts across the 3' beam in 12\,s, the CRAFTS survey of FAST benefits more from single pulse search and has discovered $\sim$31\% additional pulsars using this method. These findings indicate that there may be an extra 10-20\% of detectable pulsars which are not accounted for in our analysis.

Our limited understanding of the low end of pulsar radio luminosity function leads to a large range for the predicted pulsar yields.
The current parameters are mostly derived from Parkes Multibeam survey pulsar sample \citep[e.g.][]{YK2004,Lorimer2006,FK06,Lorimer2015}, and to build pulsar population model, it is a standard practice to calibrate the distribution parameter with PMPS sample \citep[e.g.][]{Rajwade2017,Huang2020}. So there is no surprise that the pulsar detection numbers predicted by different models and distribution parameters can converge nicely on PMPS surveys, but result in very different outcomes when applying different observing settings like FAST or FASTA which have significantly higher gain. Therefore, the result of how many new weak pulsars can be discovered in FAST and FASTA survey will provide us a very useful constraint of pulsar luminosity distribution.

\cite{Han2021} suggested the FAST pulsar surveys will probably finish with fewer new pulsar discoveries than the predicted number, indicating that the current model extrapolated from PMPS might overestimate the undetected Galactic pulsar population.
This also means that the known pulsars discovered in PMPS survey constitute a larger proportion of the overall pulsar population than we previously expected. If that is the case, some possible explanations include: (a) there are fewer pulsars at the low end of the luminosity function than we thought; (b) the distances of the pulsars estimated using their DM and the Galactic electron density model are uncertain; (c) scattering smear from the interstellar medium makes it difficult for us to detect far-away pulsars.

In conclusion, our estimation of the pulsar detection prospect of FASTA based on a variety of models generally converges to a consistent picture. Due to our limited knowledge of the real Galactic pulsar population, it is hard to pinpoint the exact number of pulsar detections of the FASTA pulsar survey with a narrow range of uncertainty. We look forward to the outcome of currently ongoing FAST pulsar surveys and future FASTA pulsar surveys with good completeness, as they will provide significantly better constraints on the Galactic pulsar population and its distribution parameters.


\normalem
\begin{acknowledgements}
This work is supported by the National Natural Science Foundation of China (NSFC) grant No. 12203070 and the National SKA Program of China No. 2020SKA0120200. This work is also supported by the NSFC grant No. 12041303, 11873067, the CAS-MPG LEGACY project. Mengyao Xue is supported by the Cultivation Project for FAST Scientific Payoff and Research Achievement of CAMS-CAS and Project funded by China Postdoctoral Science Foundation No. 2021M703237. Hongguang Wang is supported by NSFC 12133004, the National SKA Program of China No. 2020SKA0120101, the Science and Technology Program of Guangzhou (No. 202102010466) and the Astronomy Science and Technology Research Laboratory of Department of Education of Guangdong Province, China. This research made use of \texttt{PsrPopPy}\footnote{https://github.com/samb8s/PsrPopPy \citep{Bates2014}}, \texttt{PsrPopPy2}\footnote{https://github.com/devanshkv/PsrPopPy2}, and \texttt{astropy}\footnote{https://www.astropy.org \citep{Astropy2013,Astropy2018}}.
\end{acknowledgements}

\bibliographystyle{raa}
\bibliography{survey}

\end{document}